\documentclass[twocolumn,amsmath,amssymb,superscriptaddress,10pt,showkeys]
{revtex4-1}
\usepackage{color,graphicx,t1enc,grffile}

\setcitestyle{square,numbers,super}

\makeatletter
\renewcommand\NAT@citesuper[3]{\ifNAT@swa
\unskip\kern\p@\textsuperscript{\NAT@@open #1\NAT@@close}%
\if*#3*\else\ (#3)\fi\else #1\fi\endgroup}
\makeatother

\newcommand{\eqnref}[1]{Equation~(\ref{eq:#1})}
\newcommand{\picref}[1]{Figure~\ref{fig:#1}}
\newcommand{\PICREF}[1]{\textbf{Figure~\ref{fig:#1}}}
\newcommand{\chref}[1]{\ref{ch:#1}}

\newcommand{\RX}[1]{#1}

\begin{document}

\title{Simulations of Blood Flow in Plain Cylindrical and Constricted
  Vessels with Single Cell Resolution}

\author{Florian Janoschek}
\email{fjanoschek@tue.nl}
\affiliation{Department of Applied Physics, Eindhoven University of
  Technology, P.\,O.\ Box 513, 5600\,MB Eindhoven, The Netherlands}
\affiliation{Institute for Computational Physics, University of Stuttgart,
  Pfaffenwaldring 27, 70569 Stuttgart, Germany}
\author{Federico Toschi}
\email{f.toschi@tue.nl}
\affiliation{Department of Applied Physics, Eindhoven University of
  Technology, P.\,O.\ Box 513, 5600\,MB Eindhoven, The Netherlands}
\affiliation{Department of Mathematics and Computer Science, Eindhoven
  University of Technology, P.\,O.\ Box 513, 5600\,MB Eindhoven, The
  Netherlands}
\affiliation{CNR-IAC, Via dei Taurini 19, 00185 Rome, Italy}
\author{Jens Harting}
\email{j.harting@tue.nl}
\affiliation{Department of Applied Physics, Eindhoven University of
  Technology, P.\,O.\ Box 513, 5600\,MB Eindhoven, The Netherlands}
\affiliation{Institute for Computational Physics, University of Stuttgart,
  Pfaffenwaldring 27, 70569 Stuttgart, Germany}

\date{\today}

\begin{abstract}
  Understanding the physics of blood is challenging due to its nature
  as a suspension of soft particles and the fact that typical problems
  involve different scales. This is valid also for numerical
  investigations. In fact, many computational studies either neglect
  the existence of discrete cells or resolve relatively few cells very
  accurately. The authors recently developed a simple and highly
  efficient yet still particulate model with the aim to bridge the gap
  between currently applied methods. The present work focuses on its
  applicability to confined flows in vessels of diameters up to
  $\sim100\,\mu\mathrm{m}$. For hematocrit values below $30\%$, a
  dependence of the apparent viscosity on the vessel diameter in
  agreement with experimental literature data is found.
\end{abstract}
\maketitle

\section{Introduction}

Human blood can be approximated as a suspension of deformable red
blood cells (RBCs, also called erythrocytes) in a Newtonian liquid,
the blood plasma. The other constituents like leukocytes and
thrombocytes \RX{(blood platelets)} can be neglected due to their low
volume concentrations.\cite{goldsmith75} Typical concentrations for
RBCs are $40$ to $50\%$ under physiological conditions. In the absence
of external stresses, erythrocytes assume the shape of biconcave discs
of approximately $8\,\mu\mathrm{m}$ diameter.\cite{evans72} An
understanding of their effect on the rheology and the clotting
behavior of blood is necessary for the study of pathological
deviations in the body and the design of microfluidic devices for
improved blood analysis.

Well-established methods for the computer simulation of blood flow
\RX{consist of an elaborate model of deformable cell membranes coupled
  to the surrounding plasma described by particle-based hydrodynamic
  methods\cite{noguchi05,fedosov10}, the lattice Boltzmann (LB)
  method\cite{dupin07,wu10}, or the boundary integral
  method\cite{freund07}. Others} restrict themselves to a continuous
description at larger scales.\cite{boyd07} Our motivation is to bridge
the gap between both classes of models by an intermediate approach
that was published \RX{recently.}\cite{janoschek10} \RX{During the
  last decade, other groups already presented coarse-grained yet
  particulate models for blood at the mesoscale. Both Fenech et
  al.\cite{fenech09} and Chesnutt and Marshall\cite{chesnutt09}
  developed discrete element models for the aggregation of large
  numbers of RBCs in two and three dimensions. These works, however,
  do not explicitly resolve the hydrodynamics of the blood
  plasma. Approaches based on dissipative particle dynamics (DPD)
  directly model hydrodynamics. Boryczko et al.\cite{boryczko04}
  applied their model that comprises plasma, deformable red cells, and
  capillary walls to the study of fibrin aggregation in the
  microcirculation. Filipovic et al.\cite{filipovic08} employed DPD
  for the study of platelet adhesion to vessel walls and Pivkin et
  al.\cite{pivkin09} for the investigation of the effect of RBCs on
  the aggregation of blood platelets. Another simulation method that
  is well suited to account for hydrodynamics in confined geometries
  is the LB method which consequently has been applied in blood models
  of various levels of detail. Additionally to the models mentioned
  before already which are either based on a continuous
  fluid\cite{boyd07} or a suspension of fully deformable
  cells\cite{dupin07,wu10}, there exist also more coarse grained
  particulate models employing the LB method: Sun et al.\cite{sun03}
  model both red and white blood cells in two dimensions as rigid
  ellipses and spheres, the latter of which are, however, equipped
  with an elaborate wall-adhesion model. In the work by Hyakutake et
  al.\cite{hyakutake06} the behavior of two-dimensional rigid spheres
  representing RBCs at microvascular bifurcations is studied.}

\RX{Our coarse-grained blood model aims} at a minimal resolution of
red blood cells which allows for a simple and highly efficient but
still particulate description of blood as a suspension. \RX{Still, we
  do not give up explicitly modeling the suspending blood plasma or
  accounting for the non-spherical shape of RBCs.} The ultimate goal
is to perform large-scale simulations that allow to study the flow in
realistic geometries but also to link bulk properties, for example the
effective viscosity, to phenomena at the level of single
erythrocytes. Only a computationally efficient description allows the
reliable accumulation of statistical properties in time-dependent
flows which is necessary for this task. The improved understanding of
the dynamic behavior of blood might be used for the optimization of
macroscopic simulation methods.

The main idea of our model is to distinguish between the long-range
hydrodynamic coupling of cells and the short-range interactions that
are related to the complex mechanics, electrostatics, and the
chemistry of the membranes. The short-range behavior of RBCs is
described on a phenomenological level by means of anisotropic model
potentials.\cite{janoschek10} Long-range hydrodynamic interactions are
accounted for by means of \RX{an LB} method.\cite{succi01} Our model
is well suited for the implementation of complex boundary conditions
and an efficient parallelization on parallel supercomputers. Both are
necessary for the study of realistic systems like branching vessels
and the accumulation of statistically relevant data in bulk flow
situations. In Section~\chref{hydrodynamics} and~\chref{potentials},
we briefly explain our approach. For an extended description we refer
to our earlier publication.\cite{janoschek10} Section~\chref{results}
contains---next to parameter studies and a review of earlier
results\cite{janoschek10}---an investigation of the apparent blood
viscosity in tubes and the related F\r{a}hr\ae us-Lindqvist effect. It
closes with a qualitative view on constrictions and branching points
in capillaries and data regarding the scalability of the code. In
Section~\chref{conclusion}, the conclusions from our work are drawn.

\section{Hydrodynamic part of the model}\label{ch:hydrodynamics}

We apply a Bhatnagar-Gross-Krook (BGK) LB method for modeling the
blood plasma.\cite{qian92} See for example the book of Succi for a
comprehensive introduction.\cite{succi01} The single particle
distribution function $n_r(\mathbf{x},t)$ resembles the fluid
traveling with one of $r=1,\ldots,19$ discrete velocities
$\mathbf{c}_r$ at the three-dimensional lattice position $\mathbf{x}$
and discrete time $t$. Its evolution in time is determined by the
lattice Boltzmann equation
\begin{equation}
  \label{eq:lbe}
  n_r(\mathbf{x}+\mathbf{c}_r,t+1)
  =
  n_r(\mathbf{x},t)
  -
  \Omega
\end{equation}
with
\begin{equation}
  \Omega
  =
  \frac
  {n_r(\mathbf{x},t)-
    n_r^\mathrm{eq}(\varrho(\mathbf{x},t),\mathbf{u}(\mathbf{x},t))}
  {\tau}
\end{equation}
being the BGK-collision term with a single relaxation time $\tau$. The
equilibrium distribution function $n_r^\mathrm{eq}(\varrho,\mathbf{u})$ is
an expansion of the Maxwell--Boltzmann distribution.
$\varrho(\mathbf{x},t)=\sum_rn_r(\mathbf{x},t)$ and
$\varrho(\mathbf{x},t)\mathbf{u}(\mathbf{x},t)=
\sum_rn_r(\mathbf{x},t)\mathbf{c}_r$ can
be identified as density and momentum. In the limit of small
velocities and lattice spacings the Navier--Stokes equations are
recovered with a kinematic viscosity of $\nu=(2\tau-1)/6$, where
$\tau=1$ in this study.

\begin{figure}
  \centering
  \includegraphics[angle=90,width=\columnwidth]
  {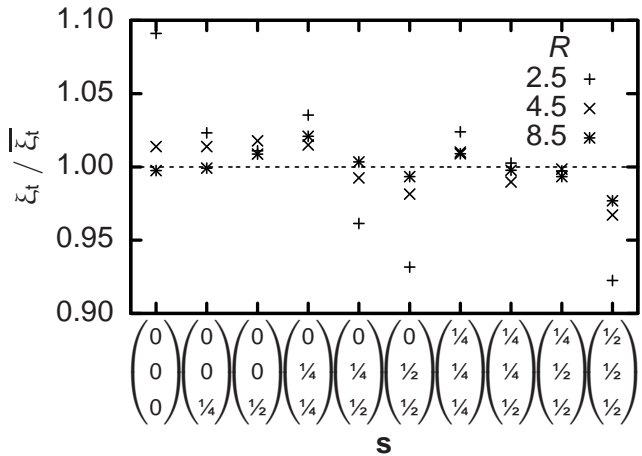}
  \caption{\RX{\label{fig:translational-discretization-error}Deviations
      of the translational drag coefficients $\xi_\mathrm{t}$ from
      their average value $\overline{\xi_\mathrm{t}}$ at selected
      offsets $\mathbf{s}$ of the particle center $\mathbf{r}_i$ to
      the nearest lattice node in the case of spherical particles with
      radius $R$.}}
\end{figure}

\begin{figure}
  \centering
  \includegraphics[angle=90,width=\columnwidth]
  {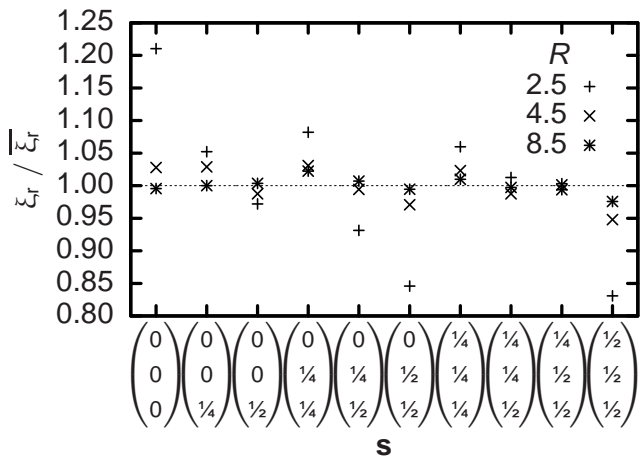}
  \caption{\RX{\label{fig:rotational-discretization-error}Deviations
      of the rotational drag coefficients $\xi_\mathrm{r}$ from their
      average value $\overline{\xi_\mathrm{r}}$ at selected offsets
      $\mathbf{s}$ of the particle center $\mathbf{r}_i$ to the
      nearest lattice node in the case of spherical particles with
      radius $R$.}}
\end{figure}

For a coarse-grained description of the hydrodynamic interaction of
cells and blood plasma, a method similar to the one by Aidun et
al. modeling rigid particles of finite size is
applied.\cite{aidun98,nguyen02} Starting point is the mid-link
bounce-back boundary condition: the confining geometry is discretized
on the lattice and all internal nodes are turned into fluid-less wall
nodes. If $\mathbf{x}$ is such a node the updated distribution at
$\mathbf{x}+\mathbf{c}_r$ is determined as
\begin{equation}\label{eq:lbe-bb}
  n_r(\mathbf{x}+\mathbf{c}_r,t+1)
  =
  n^*_{\bar{r}}(\mathbf{x}+\mathbf{c}_r,t)
  \mbox{ ,}
\end{equation}
with
\begin{equation}
  n^*_r(\mathbf{x},t)
  =
  n_r(\mathbf{x},t)
  -
  \Omega
  \;\mbox{.}
\end{equation}
This corresponds to replacing the local distribution in direction $r$
with the post-collision distribution $n^*_r(\mathbf{x},t)$ of the
opposite direction $\bar{r}$. To model boundaries moving with velocity
$\mathbf{v}$, \eqnref{lbe-bb} needs to be modified. The resulting
update rule
\begin{equation}\label{eq:lbe-bb-corr}
  n_r(\mathbf{x}+\mathbf{c}_r,t+1)
  =
  n^*_{\bar{r}}(\mathbf{x}+\mathbf{c}_r,t)
  +
  C
  \mbox{ ,}
\end{equation}
with
\begin{equation}
  C
  =
  \frac
  {2\alpha_{c_r}}
  {c_\mathrm{s}^2}
  \varrho(\mathbf{x}+\mathbf{c}_r,t)
  \,
  \mathbf{c}_r\mathbf{v}
\end{equation}
was chosen consistently with $n_r^\mathrm{eq}(\varrho,\mathbf{u})$ for the
general case $\mathbf{u}=\mathbf{v}\not=\mathbf{0}$. The lattice weights
$\alpha_{c_r}$ and the speed of sound $c_\mathrm{s}$ are constants for
a given set of discrete velocities. The momentum
\begin{equation}\label{eq:deltap}
  \Delta\mathbf{p}_\mathrm{fp}
  =
  \left(
    2n_{\bar{r}}
    +
    C
  \right)
  \mathbf{c}_{\bar{r}}
  \;\mbox{,}
\end{equation}
which is transferred during each time step by each single bounce-back
process is used to calculate the resulting force on the
boundary. \RX{A freely moving particle $i$ is modeled by such moving
  walls and is defined by its continuous position $\mathbf{r}_i$: all
  lattice nodes within a given theoretical particle volume---e.\,g. a
  sphere---centered at $\mathbf{r}_i$ are considered a wall and
  \eqnref{lbe-bb-corr} is applied to links $\mathbf{c}_r$ connected to
  its surface. When $\mathbf{r}_i$ changes, eventually the particle's
  discretization on the lattice needs to be updated.}  During this
process, fluid nodes are created or vanish and the related change in
total fluid momentum is balanced by an additional force on the
respective particle. \RX{Still, the discretized particle shape
  undergoes fluctuations depending on the offset $\mathbf{s}$ of
  $\mathbf{r}_i$ to the nearest lattice node. Their effect on
  hydrodynamics is quantified by measuring the translational and
  rotational drag coefficient $\xi_\mathrm{t}$ and $\xi_\mathrm{r}$ of
  a sphere of radius $R$ at a selection of different offset vectors
  $\mathbf{s}$. Since the model can be calibrated by assuming an
  effective hydrodynamic radius $R^*$ different from
  $R$\cite{ladd94b}, the drag coefficients are not normalized to the
  theoretical values but to their respective averages. The results in
  \textbf{Figures~\ref{fig:translational-discretization-error}} and
  \textbf{\ref{fig:rotational-discretization-error}} show that
  discretization effects depend on $R$ but even for $R=2.5$ typical
  deviations are not larger than $10\%$. Typically, deviations in the
  rotational drag coefficient appear to be two times larger than the
  corresponding translational values.}

\RX{In contrast to the biconcave equilibrium shape of physiological
  RBCs and the previous tests with spherical particles we choose a
  simplified ellipsoidal geometry that is defined by two distinct
  half-axes $R_\parallel$ and $R_\perp$ parallel and perpendicular to
  the unit vector $\hat{\mathbf{o}}_i$ which points along the
  direction of the axis of rotational symmetry of each particle $i$.}
Since the cell-fluid interaction volumes are rigid we need to allow
them to overlap in order to account for the deformability of real
erythrocytes. We assume a pair of mutual forces
\begin{equation}\label{eq:deltappp}
  \mathbf{F}_\mathrm{pp}^+
  =
  2n_r^\mathrm{eq}(\bar{\varrho},\mathbf{u}=\mathbf{0})\mathbf{c}_r
\end{equation}
and
\begin{equation}\label{eq:deltappm}
  \mathbf{F}_\mathrm{pp}^-
  =
  2n_{\bar{r}}^\mathrm{eq}(\bar{\varrho},\mathbf{u}=\mathbf{0})\mathbf{c}_{\bar{r}}
  =
  -\mathbf{F}_\mathrm{pp}^+
\end{equation}
at each cell-cell link. This is exactly the momentum transfer during
one time step due to the rigid-particle algorithm for a resting
particle and an adjacent site with resting fluid at equilibrium and
initial density $\bar{\varrho}$ and thus compensates for the static
fluid pressure. In case of close contact of cells with the confining
geometry we proceed in a similar manner as for two cells but ignore
forces on the system walls.

\section{Model potentials}\label{ch:potentials}

In order to account for the complex behavior of real RBCs at small
distances we add phenomenological pair potentials \RX{between
  cells. The idea is to develop simple model potentials and adjust
  their free parameters in order to match the pair interactions of
  cells. In a very similar way, the well-known Lennard-Jones potential
  is applied in classical molecular dynamics simulations to model
  atomic interactions. A fit can be achieved by comparison of
  simulation results and experimental data from the literature,
  specially regarding blood rheology. For the moment, we concentrate
  on high shear rates $\dot{\gamma}>10\,\mathrm{s}^{-1}$ where
  aggregative effects play no role.\cite{chien70} Therefore the model
  potential has to account only for deformation effects.} As a simple
way to describe elastic deformability, we use the repulsive branch of
a Hookean spring potential
\begin{equation}\label{eq:phi}
  \phi(r_{ij})
  =
  \left\{
    \begin{array}{l@{\qquad}l}
      \varepsilon\left(1-r_{ij}/\sigma\right)^2 &
      r_{ij}<\sigma\\
      0 &
      r_{ij}\ge\sigma
    \end{array}
  \right.
\end{equation}
for the scalar displacement $r_{ij}$ of two cells $i$ and $j$. With
respect to the disc-shape of RBCs, we follow the approach of Berne and
Pechukas\cite{berne72} and choose the energy and range parameters
\begin{equation}\label{eq:epsilon}
  \varepsilon(\hat{\mathbf{o}}_i,\hat{\mathbf{o}}_j)
  =
  \frac
  {\bar{\varepsilon}}
  {\sqrt{1-\chi^2\left(\hat{\mathbf{o}}_i\hat{\mathbf{o}}_j\right)^2}}
\end{equation}
and
\begin{equation}\label{eq:sigma}
  \sigma(\hat{\mathbf{o}}_i,\hat{\mathbf{o}}_j,\hat{\mathbf{r}}_{ij})
  =
  \frac
  {\bar{\sigma}}
  {\sqrt{1-\frac{\chi}{2}\left[
        \frac
        {\left(
            \hat{\mathbf{r}}_{ij}\hat{\mathbf{o}}_i
            +
            \hat{\mathbf{r}}_{ij}\hat{\mathbf{o}}_j
          \right)^2}
        {1+\chi\hat{\mathbf{o}}_i\hat{\mathbf{o}}_j}
        +
        \frac
        {\left(
            \hat{\mathbf{r}}_{ij}\hat{\mathbf{o}}_i
            -
            \hat{\mathbf{r}}_{ij}\hat{\mathbf{o}}_j
          \right)^2}
        {1-\chi\hat{\mathbf{o}}_i\hat{\mathbf{o}}_j}
      \right]}}
\end{equation}
as functions of the orientations $\hat{\mathbf{o}}_i$ and
$\hat{\mathbf{o}}_j$ of the cells and their normalized center
displacement $\hat{\mathbf{r}}_{ij}$. We achieve an anisotropic
potential with a zero-energy surface that is approximately that of
ellipsoidal discs. Their half-axes parallel $\sigma_\parallel$ and
perpendicular $\sigma_\perp$ to the symmetry axis enter
\eqnref{epsilon} and~(\ref{eq:sigma}) via $\bar{\sigma}=2\sigma_\perp$
and
$\chi=(\sigma_\parallel^2-\sigma_\perp^2)/(\sigma_\parallel^2+\sigma_\perp^2)$
whereas $\bar{\varepsilon}$ determines the potential strength. For
modeling the cell-wall interaction we assume a sphere with radius
$\sigma_\mathrm{w}=1/2$ at every lattice node on the surface of a
vessel wall and implement similar forces as for the cell-cell
interaction. Using
\begin{equation}\label{eq:sigmaw}
  \sigma(\hat{\mathbf{o}}_i,\hat{\mathbf{r}}_{i\mathbf{x}})
  =
  \frac
  {\bar{\sigma}_\mathrm{w}}
  {\sqrt
    {1
      -
      \chi_\mathrm{w}
      \left(\hat{\mathbf{r}}_{i\mathbf{x}}\hat{\mathbf{o}}_i\right)^2}}
\end{equation}
as a range parameter with
$\bar{\sigma}_\mathrm{w}=\sqrt{\sigma_\perp^2+\sigma_\mathrm{w}^2}$
and $\chi_\mathrm{w}=
(\sigma_\parallel^2-\sigma_\perp^2)/(\sigma_\parallel^2+\sigma_\mathrm{w}^2)$
allows to scale a potential with radial symmetry to fit for the
description of the interaction of a sphere and an ellipsoidal
disc.\cite{berne72} The parameter
$\varepsilon(\hat{\mathbf{o}}_i,\hat{\mathbf{o}}_j)=\bar{\varepsilon}_\mathrm{w}$
remains constant in this case. $\hat{\mathbf{r}}_{i\mathbf{x}}$ is the
normalized center displacement of cell $i$ and a wall node
$\mathbf{x}$.

\begin{figure}
  \centering
  \includegraphics[width=\columnwidth]{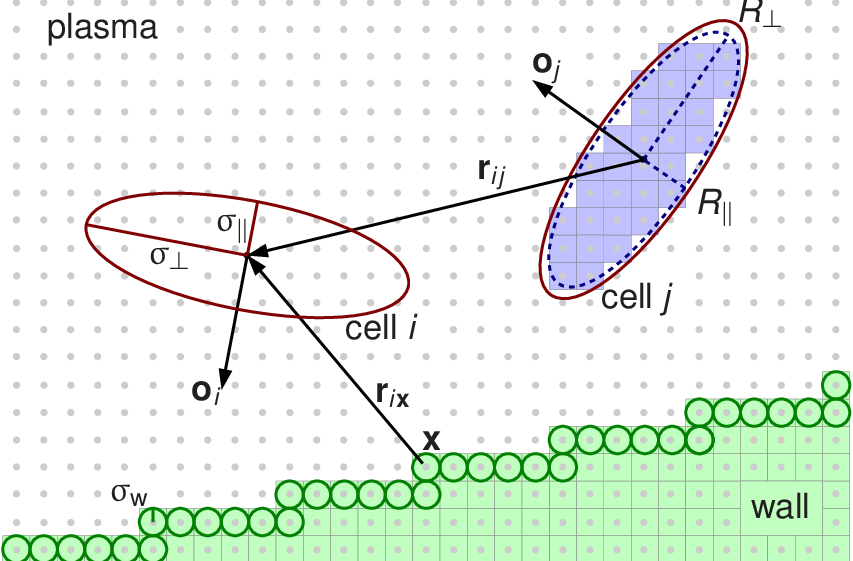}
  \caption{\label{fig:model-2d}Two-dimensional cut as an outline of
    the model. Shown are two cells with their axes of rotational
    symmetry depicted by $\hat{\mathbf{o}}_{i/j}$. The volumes defined
    by the cell-cell interaction are approximately ellipsoidal with
    half-axes $\sigma_{\perp/\parallel}$. The smaller ellipsoidal
    volumes (half-axes $R_{\perp/\parallel}$) of the cell-plasma
    interaction are discretized on the underlying lattice. The
    cell-wall potential assumes spheres with radius
    $\sigma_\mathrm{w}$ on all surface wall nodes $\mathbf{x}$ (taken
    from Reference\cite{janoschek10}).}
\end{figure}

\PICREF{model-2d} shows a two-dimensional cut as an outline of the
full model. For two RBCs the inner volume implementing the cell-plasma
interaction with half axes $R_\parallel$ and $R_\perp$ is shown. Also
depicted are the larger volumes that are defined by the range
parameters $\sigma_\parallel$ and $\sigma_\perp$ of the cell-cell and
cell-wall interaction. A section of a vessel wall is visualized. The
cell-wall interaction is based on the assumption of spheres with
radius $\sigma_\mathrm{w}$ at the location of the surface wall
nodes. The forces \RX{and torques} emerging from the interaction of
the cells with the fluid, other RBCs, and walls are integrated by a
\RX{leap-frog scheme} in order to evolve the system in time. \RX{Both
  the LB routines and the ones employed for treatment of the suspended
  cells use the same domain decomposition.} For more detailed
information concerning the model we refer to.\cite{janoschek10}

\section{Results}\label{ch:results}

All quantities can be converted from simulation units to physical
units by multiplication with products of integer powers of the
conversion factors $\delta x$, $\delta t$, and $\delta m$ for space,
time, and mass. As a convention in this work, primed variables are
used to distinguish quantities given in physical units from the same
unprimed variable measured in lattice units. Based on experimental
measurements of RBC geometries,\cite{evans72} the half-axes of the
simplified volume defining the cell-cell interaction in our model are
set to
\begin{equation}\label{eq:sigmasp}
  \sigma'_\perp=4\,\mu\mathrm{m}
  \quad\mbox{and}\quad
  \sigma'_\parallel=4/3\,\mu\mathrm{m}
  \mbox{ .}
\end{equation}
When choosing the spatial resolution quantified by the physical
distance $\delta x$ corresponding to one lattice spacing, a compromise
needs to be found: A low resolution severely reduces the computational
effort but also reduces numerical accuracy. We decide for $\delta
x=2/3\,\mu\mathrm{m}$\cite{janoschek10} which allows a contiguous and
closed volume of the cell-fluid interaction which still is completely
comprised within the cell-cell interaction volume~\eqnref{sigmasp}. On
the actual values of $R_\perp$ and $R_\parallel$ we decide after
studying their influence on the apparent viscosity $\mu_\mathrm{app}$
of the model suspension in plane Couette flow.

Supposing that $\nu$ matches the kinematic plasma viscosity of
$\nu'=1.09\times10^{-6}\,\mathrm{m}^2\cdot\mathrm{s}^{-1}$, the time
discretization is determined as $\delta
t=6.80\times10^{-8}\,\mathrm{s}$. \RX{Since the largest shear rates
  employed in this work are of the order of $10^4\,\mathrm{s}^{-1}$,
  sufficient temporal resolution is provided.} $\delta
m=3.05\times10^{-16}\,\mathrm{kg}$ is chosen arbitrarily. All shear
flow simulations reported here are performed on a system with a size
of $l_x=128$ lattice units in $x$- and $l_y=l_z=40$ lattice units in
$y$- and $z$-direction or $85\times27^2\,\mu\mathrm{m}^3$ of real
blood. Between the two $yz$-side planes a constant offset of the local
fluid velocities in $z$-direction is imposed by an adaption of the
Lees-Edwards shear boundary condition to the LB method.\cite{wagner99}
The apparent viscosity is calculated from the average shear rate and
shear stress.\cite{janoschek10} Recently, an alternative method of
viscosity measurement based on Kolmogorov flow was demonstrated which
allows the employment of more simple periodic boundary
conditions.\cite{janoschek10b}

The influence of the volume of the cell-fluid interaction is studied
as a function of the shear rate. \PICREF{shear-ladd-shape} shows the
result for different $R_\parallel$ and $R_\perp$ at a cell number
concentration that varies by less than $5\%$. Generally, shear
thinning is visible, but both the absolute viscosities and the change
in viscosity per shear rate increase significantly for larger
interaction volumes. Based on this and further parameter
studies,\cite{janoschek10} we choose with
$R'_\perp=11/3\,\mu\mathrm{m}$ the largest value investigated here and
assuming $R_\parallel/\sigma_\parallel=R_\perp/\sigma_\perp=11/12$
obtain $R'_\parallel=R'_\perp/3=11/9\,\mu\mathrm{m}$\RX{\ or
  $R_\parallel=1.833$ and $R_\perp=5.5$ measured in lattice units. A
  sphere with equal volume has a radius of
  $R=3.813$. Figures~\ref{fig:translational-discretization-error} and
  \ref{fig:rotational-discretization-error} suggest that at this
  resolution, we have to expect discretization errors of the order of
  some percents that are acceptable for our approach}.

\begin{figure}
  \centering
  \includegraphics[width=\columnwidth]{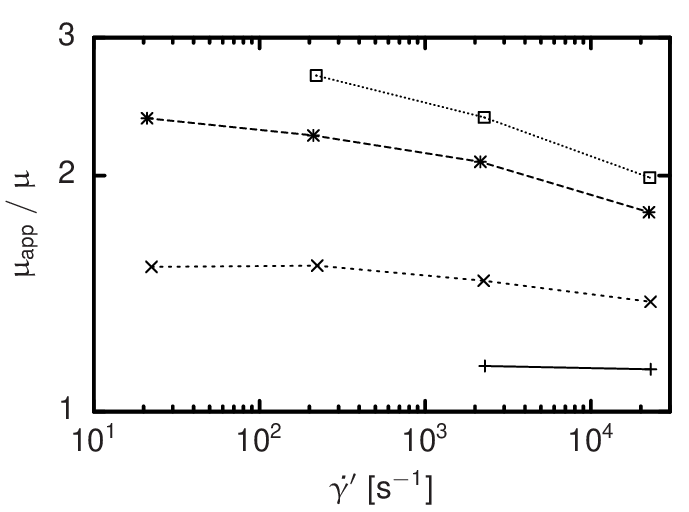}
  \caption{\label{fig:shear-ladd-shape}Apparent viscosity
    $\mu_\mathrm{app}$ in dependence on the shear rate $\dot{\gamma}$
    as calculated for Couette flow for different sets of model
    parameters. $\sigma_\perp'=4\sigma_\parallel'=4\,\mu\mathrm{m}$
    and $\bar{\varepsilon}'=1.47\times10^{-14}\,\mathrm{J}$ are kept
    fixed. While the volume implementing the cell-fluid interaction is
    varied from bottom to top as
    $R'_\perp=R'_\parallel=1\,\mu\mathrm{m}$,
    $R'_\perp=4R'_\parallel=8/3\,\mu\mathrm{m}$,
    $R'_\perp=4R'_\parallel=10/3\,\mu\mathrm{m}$, and
    $R'_\perp=4R'_\parallel=11/3\,\mu\mathrm{m}$, the cell-fluid
    volume concentration $\Phi$ varies within $2.6\%$ and
    $36\%$. Larger volumes $4\pi R_\perp^2R_\parallel/3$ lead to more
    pronounced shear thinning and generally higher viscosities.}
\end{figure}

In further simulations the effect of the stiffness parameter of the
cell-cell potential $\bar{\varepsilon}$ is studied. The viscosity as a
function of the shear rate increases with increasing
$\bar{\varepsilon}$. For very stiff cells this dependence on the shear
rate decreases considerably which is in asymptotic consistency with
the experimental results of Chien\cite{chien70} who measured the
apparent shear viscosity of a suspension of artificially hardened RBCs
and found a significantly increased yet mostly constant viscosity. At
a cell-fluid volume concentration of $43\%$ which seems sufficiently
close to the hematocrit of $45\%$ in the measurements done by
Chien\cite{chien70} to justify a quantitative comparison, we find best
agreement for $\bar{\varepsilon}'=1.47\times10^{-15}\,\mathrm{J}$ and
use this parametrization for all following
investigations.\cite{janoschek10}

We now confine our model suspension in a cylindrical channel of
diameter $D$ and a length of $43\,\mu\mathrm{m}$ with periodic
boundaries. Both cells and plasma are steadily driven by a volume
force equivalent to a pressure gradient $\mathrm{d}P/\mathrm{d}z$. We
arbitrarily choose
$\bar{\varepsilon}'_\mathrm{w}=1.47\times10^{-16}\,\mathrm{J}$ for the
strength of the cell-wall interaction as a value that reliably
prevents cells from penetrating the vessel
wall. \PICREF{channel-t2b-r047-n0.6-slow} visualizes the cells as the
volumes defined by the cell-cell interaction and the channel wall as
midplane between fluid and wall nodes in the case of
$D'=63\,\mu\mathrm{m}$ and $\Phi=25\%$. The pseudo-shear rate, which
is defined via the volume flow rate $Q$, is $\bar{v}'=4Q'/(\pi
D'^3)=61\,\mathrm{s}^{-1}$. A preferential alignment of the cells
largely perpendicular to the velocity gradient is visible.

\begin{figure}
  \centering
  \includegraphics[width=0.8\columnwidth]{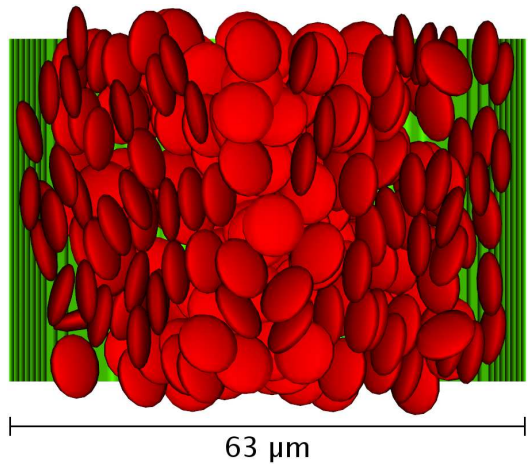}
  \caption{\label{fig:channel-t2b-r047-n0.6-slow}Steady flow through a
    cylindrical channel with a diameter of $D'=63\,\mu\mathrm{m}$ at
    $\Phi=25\%$ cell-fluid volume concentration. The volumes defined
    by the cell-cell interaction are displayed at a cut parallel to
    the center axis. The flow is pointing from top to bottom and
    effects a pseudo-shear rate of
    $\bar{v}'=61\,\mathrm{s}^{-1}$. Alignment of cells with the shear
    flow is observed.}
\end{figure}

We compare radial velocity profiles for different flow velocities, a
cell-fluid volume concentration of $\Phi=42\%$, and
$D'=63\,\mu\mathrm{m}$. While for high velocities the result looks
parabolic in the central region, there is increasing blunting of the
profile when the flow rate is reduced. The blunting can be understood
as a consequence of the shear-thinning behavior of the model and is
qualitatively consistent with experimental data from the
literature.\cite{goldsmith75} Generally, apparent slip is visible
close to the vessel wall. It is due to a cell depletion layer that to
some extent can be controlled via
$\bar{\varepsilon}_\mathrm{w}$. Thus, the observations can partly be
described by an existing model assuming a homogeneous core region with
high hematocrit and consequently high viscosity and a cell-depleted
boundary layer close to the vessel wall.\cite{janoschek10,popel05}

\begin{figure}
  \centering
  \includegraphics[width=\columnwidth]{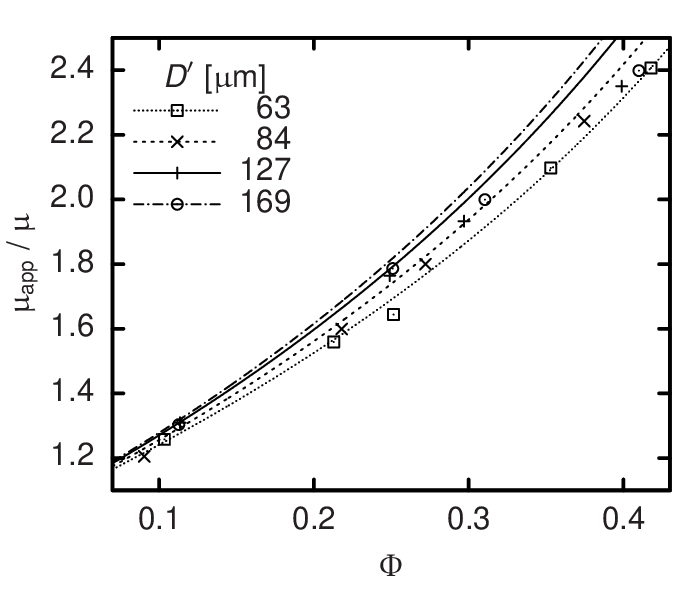}
  \caption{\label{fig:channel-n-viscosity-medium}Same plot as in
    \picref{channel-n-viscosity-slow} but for
    $\bar{v}'=(180\pm5)\,\mathrm{s}^{-1}$ and two additional vessel
    diameters $D$. Due to shear-thinning, the viscosities here are
    slightly lower in general. For $\Phi\lesssim0.3$, the dependence
    of $\mu_\mathrm{app}$ on $D$ is captured again, but only on a more
    qualitative level.}
\end{figure}

As it is well known, the formation of this cell depletion layer
influences the flow resistance of vessels which can be expressed in
terms of their apparent viscosity.\cite{popel05} By inserting the
measured volume flow rate $Q$ through a cylindrical vessel into the
theoretical expression for a Newtonian fluid
\begin{equation}\label{eq:hp}
  Q
  =
  \frac
  {\pi D^4}
  {128\mu}
  \frac{\mathrm{d}P}{\mathrm{d}z}
\end{equation}
and solving for $\mu$, the respective apparent viscosity
$\mu_\mathrm{app}$ can be determined. \RX{Except for $Q$, all known
  quantities in \eqnref{hp} are constant and set as parameters of our
  simulation code. Only the flow rate undergoes stochastic
  fluctuations and---at the beginning of each simulation---shows a
  strong time dependence as the system relaxes from an arbitrary
  initial condition to a macroscopically steady state. Thus, typical
  simulations last for up to $\sim10^7$ time steps corresponding to
  $\sim 1\,\mathrm{s}$ of physical time. $Q$ is averaged over
  typically $\sim0.1\,\mathrm{s}$ and its statistical error serves to
  estimate the accuracy of the resulting viscosity. The main
  dependency of $\mu_\mathrm{app}$ is on the hematocrit $\Phi$. A
  comparison with experimental data in the case of Couette flow was
  published earlier.\cite{janoschek10} However, $\mu_\mathrm{app}$
  depends also on the vessel diameter $D$.} This is known as
F\r{a}hr\ae us-Lindqvist effect.\cite{popel05} Pries et al.\ combine a
large set of experimental studies for $\bar{v}'>50\,\mathrm{s}^{-1}$
and provide an empirical fit.\cite{pries92b} The resulting expression
for $\mu_\mathrm{app}$ is plotted as a function of $\Phi$ for four
discrete values of $D$ as lines in \PICREF{channel-n-viscosity-slow}
and \textbf{\ref{fig:channel-n-viscosity-medium}}. The symbols stand
for simulation results at the same diameters $D$ and pseudo-shear
rates of $(62\pm1)\,\mathrm{s}^{-1}$
(\picref{channel-n-viscosity-slow}) and $(180\pm5)\,\mathrm{s}^{-1}$
(\picref{channel-n-viscosity-medium}). In consistency with the still
significant shear-thinning that experiments\cite{chien70} but also our
simulations\cite{janoschek10} exhibit above
$\dot{\gamma}'=50\,\mathrm{s}^{-1}$, our results show a clear
dependency on the pseudo-shear rate, which cannot be covered by the
curves by Pries et al.\ that were obtained from averaging over
different $\bar{v}'>50\,\mathrm{s}^{-1}$.\cite{pries92b} Nevertheless,
a comparison confirms the agreement concerning the order of magnitude
of our results and the observation that the apparent viscosity
increases with $D$. This can be explained by the decreasing relative
influence of the cell depletion layer. While the effect seems captured
realistically for low volume concentrations $\Phi\lesssim0.3$, the
dependence of $\mu_\mathrm{app}$ on $D$ becomes less clear for higher
$\Phi$. We explain this discrepancy by the fact that the thickness of
the cell-depleted layer at high $\Phi$ is determined by a balance of
the short-range interactions of cells in the core acting towards a
decrease of the depletion layer thickness and the short-range
interactions of cells and the vessel wall acting towards its
increase. Since the method presented above models short-range
interactions by means of potential forces and a constant pressure
force~\eqnref{deltappp} and~(\ref{eq:deltappm}), velocity-dependent
lubrication and lift forces can be covered only insufficiently in the
case of very high volume concentrations. However, due to the
F\r{a}hr\ae us effect,\cite{popel05} $\Phi$ in smaller blood vessels
is lower than the discharge hematocrit of typically $40$--$50\%$. Also
the $D$-dependence found according to Pries et al.\cite{pries92b} and
displayed in \picref{channel-n-viscosity-slow}
and~\ref{fig:channel-n-viscosity-medium} is not very strong for the
diameters studied. Thus the actual impact of the limitation of our
model concerning $\Phi$ should not be overestimated when simulating
vessels of comparable diameters.

\begin{figure}
  \centering
  \includegraphics[width=\columnwidth]{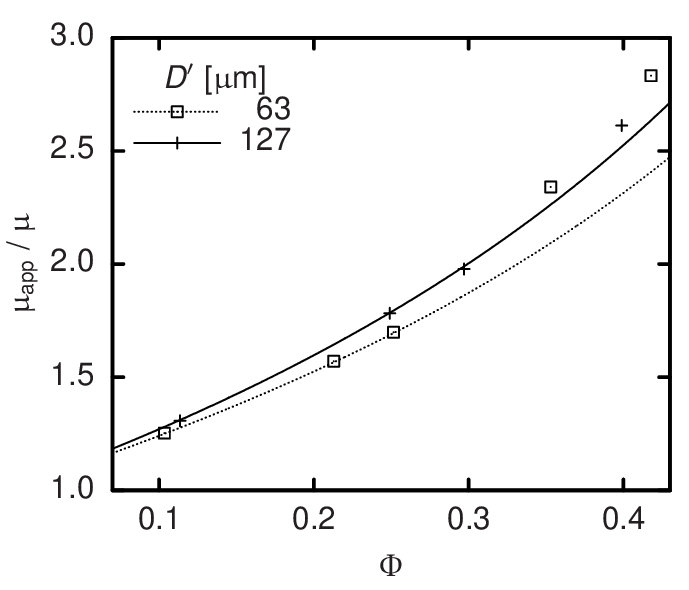}
  \caption{\label{fig:channel-n-viscosity-slow}Dependence of the
    apparent viscosity $\mu_\mathrm{app}$ in a cylindrical vessel with
    diameter $D$ on the volume concentration $\Phi$. The lines show
    empirical results by Pries et al.\cite{pries92b} for pseudo-shear
    rates $\bar{v}'>50\,\mathrm{s}^{-1}$ with $\Phi$ as tube
    hematocrit while the symbols represent our simulations for
    $\bar{v}'=(62\pm1)\,\mathrm{s}^{-1}$. Error bars are of the order
    of the symbol diameters and thus are not drawn. For
    $\Phi\lesssim0.3$ there is good consistency of simulation results
    and experimental data.}
\end{figure}

With a concluding study of the flow of nine RBCs through branching
capillaries we demonstrate on a qualitative level that our
coarse-grained model is able to describe problems where even single
cells significantly affect the whole flow. While the diameter of the
capillaries amounts to $9.3\,\mu\mathrm{m}$, one of the branches
features a stenosis of only $5.3\,\mu\mathrm{m}$. A cut of this setup
is visualized in \PICREF{streamlines-lowere}. Both tube diameters and
Reynolds numbers $Re\lesssim4\times10^{-3}$ match physiological
situations.\cite{fung81} We vary the cell-wall interaction stiffness
$\bar{\varepsilon}_\mathrm{w}$ and find that for
$\bar{\varepsilon}'_\mathrm{w}=1.47\times10^{-17}\,\mathrm{J}$, the
erythrocytes easily pass the constriction as expected for healthy
cells.\cite{goldsmith75} The trajectories of the cell centers in this
case are also shown in \picref{streamlines-lowere}. They visualize
that---as expected from the literature\cite{fung81}---a clear majority
of cells is drawn into the unconstricted branch due to its higher flow
rate. In consequence, the branch with the stenosis features a reduced
hematocrit. \PICREF{flow-tiny-r4}, which displays the time evolution
of the relative volume flow rate in the constricted branch, makes
clear that for
$\bar{\varepsilon}'_\mathrm{w}=1.47\times10^{-17}\,\mathrm{J}$ this
quantity changes only due to fluctuations induced by the mutable
configuration of RBCs in the system. The constricted branch gets
clogged for
$\bar{\varepsilon}'_\mathrm{w}=1.47\times10^{-15}\,\mathrm{J}$ which
is also reflected by \picref{flow-tiny-r4}. In the case of an
intermediate value of
$\bar{\varepsilon}'_\mathrm{w}=1.47\times10^{-16}\,\mathrm{J}$, cells
are initially stopped but due to pressure fluctuations get squeezed
through the stenosis eventually. The acceleration by the cell-wall
potential when leaving the constriction leads to the peaks of the flow
rate visible in \picref{flow-tiny-r4}.

\begin{figure}
  \centering
  \includegraphics[angle=90,width=\columnwidth]{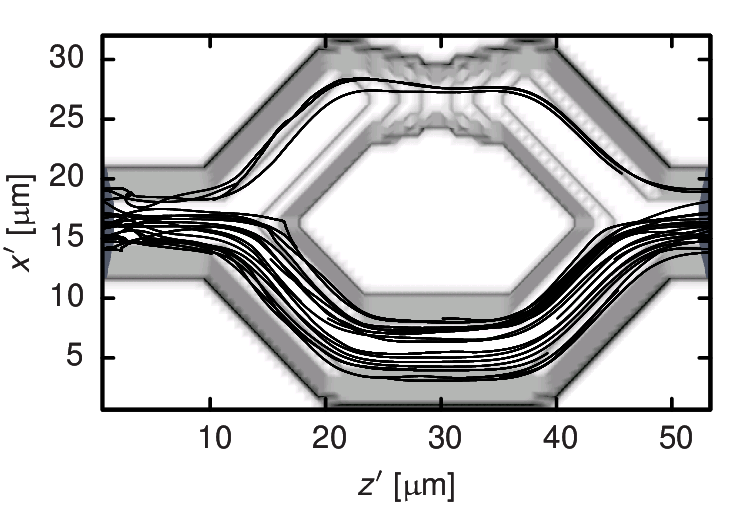}
  \caption{\label{fig:streamlines-lowere}Branching capillaries
    visualized as a cut of the midplane between wall and fluid
    nodes. The channel diameter is $5.3\,\mu\mathrm{m}$ at the
    stenosis in the upper branch and $9.3\,\mu\mathrm{m}$
    otherwise. The flow direction is pointing from left to right. The
    plot also shows cell center trajectories at a cell-wall
    interaction stiffness of
    $\bar{\varepsilon}'_\mathrm{w}=1.47\times10^{-17}\,\mathrm{J}$.}
\end{figure}

\begin{figure}
  \centering
  \includegraphics[width=\columnwidth]{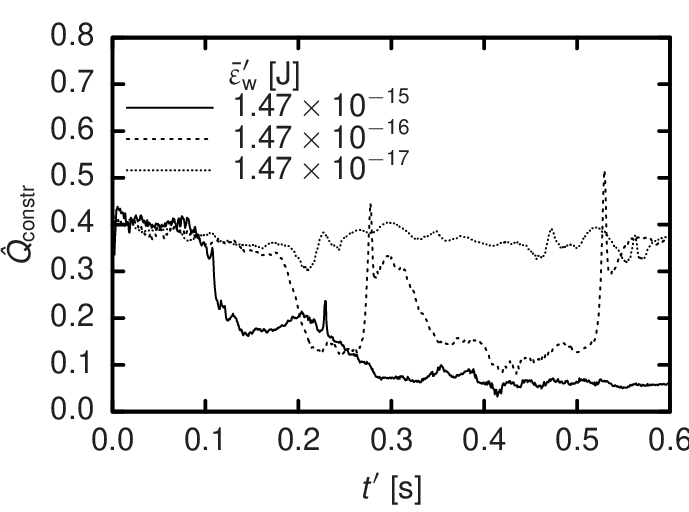}
  \caption{\label{fig:flow-tiny-r4}Time evolution of the relative
    volume flow rate in the constricted branch of the system depicted
    in \picref{streamlines-lowere} at different values of
    $\bar{\varepsilon}'_\mathrm{w}$. The clogging in the case of
    $\bar{\varepsilon}'_\mathrm{w}=1.47\times10^{-15}\,\mathrm{J}$
    becomes visible in a drop of the relative flow rate to less than
    $10\%$. While
    $\bar{\varepsilon}'_\mathrm{w}=1.47\times10^{-17}\,\mathrm{J}$
    leads to a continuous flow situation (see
    \picref{streamlines-lowere}), there are temporary drops and sharp
    peaks of the flow rate for
    $\bar{\varepsilon}'_\mathrm{w}=1.47\times10^{-16}\,\mathrm{J}$. They
    can be explained by RBCs being initially stopped at the stenosis
    and eventually squeezed through due to local pressure fluctuations
    (taken from Reference\cite{janoschek10}).}
\end{figure}

While the results for the straight channel studied first could still
mostly be reproduced by a homogenous fluid with a specially tuned
shear-rate or position dependent viscosity,\cite{boyd07,secomb03} this
is not possible at the scale of capillaries. Here, our model allows to
account for clearly particulate effects like clogging or local changes
of flow rate and pressure. In \picref{flow-tiny-r4}, it is
demonstrated that such effects can lead to a distinct unsteadiness of
the local cell volume concentration which is present also in human
microvascular networks.\cite{popel05}

Despite the simplifications of the model, parallel supercomputers are
necessary to simulate more realistic vessel networks or large bulk
systems. This makes the scalability of the code crucial. For a
quasi-homogenous chunk of suspension consisting of $1024^2\times2048$
lattice sites and $4.1\times10^6$ cells (see \PICREF{huge}) simulated
on a BlueGene/P system, we achieve a parallel efficiency normalized to
the case of $2048$-fold parallelism of $85.6\%$ on $32768$ and still
$67.3\%$ on $65536$ cores. In comparison, the pure LB code without
\RX{suspended RBCs} shows a relative parallel efficiency of $98.0\%$
on $65536$ cores. The parallel performance of the combined code is
mainly limited by the relation of the interaction range of a cell to
the size of the computational domain dedicated to each task. The full
strong scaling behavior for up to $65536$ cores is shown in
\PICREF{speedup-jugene-big}. Compared to deformable particle
models,\cite{dupin07} our method not only has a lower overall number
of computations at a given resolution but is also easier to
parallelize efficiently because each RBC has only six degrees of
freedom.

\begin{figure}
  \centering
  \includegraphics[width=0.8\columnwidth]{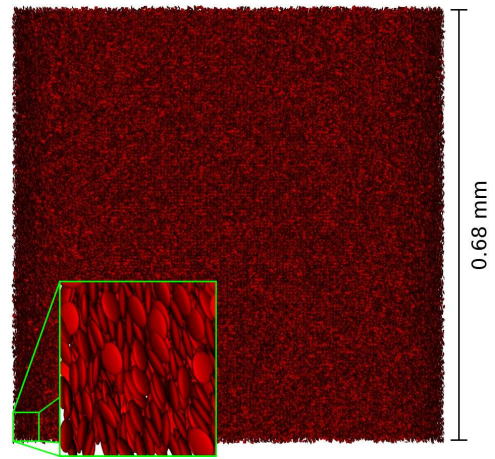}
  \caption{\label{fig:huge}Schematic view of one of the square side
    planes of a benchmark system containing $1024^2\times2048$ lattice
    sites and more than $4\times10^6$ RBCs. The simulated volume
    resembles $0.68^2\times1.37\,\mathrm{mm}^3$ of blood.}
\end{figure}

\begin{figure}
  \includegraphics[width=\columnwidth]{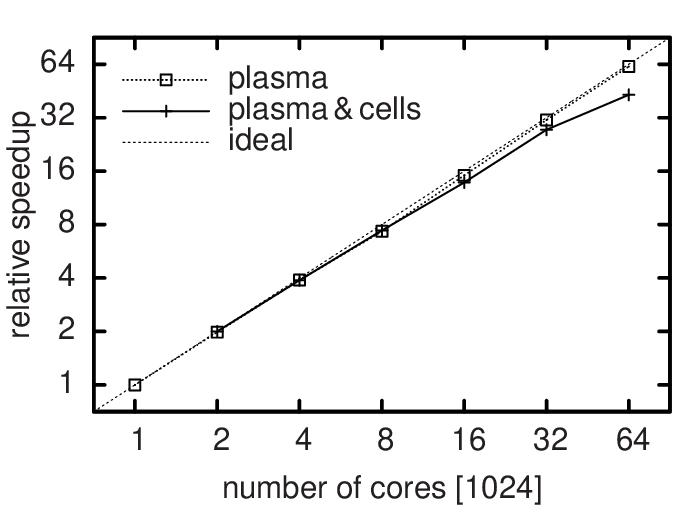}
  \caption{\label{fig:speedup-jugene-big}\RX{Strong scaling benchmark
      for the system depicted in \picref{huge} simulated on the
      BlueGene/P system at JSC. The relative speedup normalized to
      $1024$ cores is plotted as a function of the number of cores for
      the full model (plasma\,\&\,cells) and a cell-free fluid volume
      of the same size (plasma). Due to higher memory requirements,
      the tasks in the combined case were performed by at least $2048$
      cores.}}
\end{figure}

\section{Conclusion}\label{ch:conclusion}

We recently developed a computational model for the coarse-grained
simulation of blood flow.\cite{janoschek10} In the present work,
additional bulk flow studies regarding the model parameters are
shown. We further extend previous measurements of the apparent
viscosity in cylindrical channels to a comparison of different tube
diameters between $63$ and $169\,\mu\mathrm{m}$. While at higher
hematocrit values the consistency remains limited to capturing the
general behavior of the viscosity and its order of magnitude, the
F\r{a}hr\ae us-Lindqvist effect is reproduced very similarly to
experimental data for $\Phi\lesssim0.3$.\cite{pries92b} With a
qualitative study of unsteady flow and clogging in branching
capillaries with a constriction, possible applications of our model
are shown to range from bulk flow to capillary networks. Taking into
account the good scalability of the code on massively parallel
supercomputers, we believe that our method can help to achieve a
better understanding of the links between the behavior of single cells
and the properties of the encompassing larger flow structures.

\begin{acknowledgments}
  The authors acknowledge financial support from the Netherlands
  Organization for Scientific Research (VIDI grant of J.~Harting), the
  TU/e High Potential Research Program, and the HPC-Europa2 project as
  well as computing resources from JSC J\"ulich (granted by PRACE),
  SSC Karlsruhe, CSC Espoo, EPCC Edinburgh, and SARA Amsterdam, the
  latter three being granted by DEISA as part of the DECI-5 project.
\end{acknowledgments}

%

\section{Material for table of contents}

\subsection{Picture}

\includegraphics[width=5.5cm]{channel-t2b-r047-n0.6-slow}

\subsection{Text}

\textbf{The simulation of hemodynamics is a very challenging task.}
Recently, a scalable and simple particulate model was developed that
has the potential to bridge the gap that existing methods left
open. Here, the model is applied to flows in vessels of diameters up
to $\sim100\,\mu\mathrm{m}$.

\end{document}